\documentclass{ws-p8-50x6-00}

\begin{document}

\thispagestyle{empty}
\noindent
\begin{flushright}
        OHSTPY-HEP-T-01-025 \\
                August 2001
\end{flushright}

\vspace{1cm}
\begin{center}
  \begin{Large}
  \begin{bf}
Yukawa Coupling Unification in SO(10) Supersymmetric Grand
Unified Theories\footnote{Talk given at the 9th International 
Conference on Supersymmetry and Unification of Fundamental 
Interactions SUSY'01, June 11--17, Dubna, Russia.}\footnote{This talk is based on the 
work done in collaboration with T. Bla\v zek and S. Raby.}\footnote{This work was 
supported in part by DOE grant DOE/ER/01545-820.}
\end{bf}                  
 \end{Large}              
\end{center}
  \vspace{1cm}
    \begin{center}
Radovan Derm\' \i \v sek \\
      \vspace{0.3cm}
\begin{it}
Department of Physics, The Ohio State University, \\
174 W. 18th Ave., Columbus, Ohio  43210
\end{it}
  \end{center}
  \vspace{1cm}
\centerline{\bf Abstract}
\begin{quotation}
\noindent   
We report on a preferred region of supersymmetric
parameter space by $t,\; b,\;\tau$ Yukawa unification.
We find a narrow region survives for $\mu > 0$ (suggested by   
$b\rightarrow s\gamma$ and the anomalous magnetic moment of the
muon) with  $A_0 \sim - 2 \, m_{16}$, $m_{16}
> 1200$ GeV, $\mu, M_{1/2} \ll  \, m_{16}$ and $\tan \beta \sim 50
\pm 2$. Demanding Yukawa unification thus makes definite
predictions for Higgs and sparticle masses.
\end{quotation}   

\newpage

\title{Yukawa Coupling Unification in SO(10) Supersymmetric Grand
Unified Theories}

\author{Radovan Derm\'i\v sek}

\address{Department of Physics, The Ohio State University, \\
174 W. 18th Ave., Columbus, OH 43210, USA \\
E-mail: dermisek@pacific.mps.ohio-state.edu}

%%%%%%%%%%%%%%%%%%%%%%%%%%%%%%%%%%%%%%%%%%%%%%%%%%%%%%%%%%%%%%
% You may repeat \author \address as often as necessary      %
%%%%%%%%%%%%%%%%%%%%%%%%%%%%%%%%%%%%%%%%%%%%%%%%%%%%%%%%%%%%%%

\maketitle

\abstracts{ We report on a preferred region of supersymmetric
parameter space by $t,\; b,\;\tau$ Yukawa unification.
We find a narrow region survives for $\mu > 0$ (suggested by
$b\rightarrow s\gamma$ and the anomalous magnetic moment of the
muon) with  $A_0 \sim - 2 \, m_{16}$, $m_{16}
> 1200$ GeV, $\mu, M_{1/2} \ll  \, m_{16}$ and $\tan \beta \sim 50
\pm 2$. Demanding Yukawa unification thus makes definite
predictions for Higgs and sparticle masses. }

\noindent The nicest features of supersymmetric [SUSY] grand
unified theories [GUTs] are the explanation of charge assignments
of quarks and leptons and the prediction of the weak mixing angle
from the unification of gauge couplings at the GUT scale $M_G \sim 3
\times 10^{16}$ GeV. Furthermore, there is also a very interesting
possibility that the Yukawa couplings of the third generation
particles unify at the same scale. The unification of Yukawa
couplings of $t,\; b,\;\tau$ is very well motivated in $SO(10)$
models where the third generation standard model [SM] fermions and
the righthanded neutrino sit in one {\bf 16} dimensional spinor
representation and (in the simplest version) the two Higgs
doublets of minimal supersymmetric standard model are
contained in one {\bf 10} dimensional fundamental representation.
The Yukawa couplings are given by a single term in the
superpotential $W = \lambda \, \bf \, 16 \, 10 \, 16$ which leads 
to 
$\lambda_t = \lambda_b = \lambda_\tau = \lambda$ at the GUT scale.
Besides $\lambda$ and $M_G$ there is one more parameter crucial
for fermion masses -- $\tan \beta$, the ratio of vevs of two
Higgses. Since $M_G$ is already fixed by gauge coupling
unification there are just two parameters to play with and it is
non-trivial that one can describe three masses by one Yukawa
coupling. 
%All other parameters: gaugino, squark and slepton masses
%affect masses of quarks and leptons through SUSY threshold
%corrections and thus whether Yukawa unification works or not is
%connected with the way SUSY is broken. 
In what follows we specify
the SUSY breaking scenario and comment on alternatives, we show
that there is a preferred region of SUSY parameter space where
Yukawa unification works, and give predictions for Higgs and SUSY
spectra from this specific region. For a more detailed discussion
see Ref.\cite{BDR}.

In our analysis we assume following boundary conditions 
for soft SUSY breaking terms at $M_G$: universal squark 
and slepton masses
-- $m_{16}$; universal gaugino masses -- $M_{1/2}$; non-universal
Higgs $H_u, \, H_d$ masses; and universal trilinear couplings --
$A$. The special case $m_{16} = m_{H_u} = m_{H_d}$ corresponds to the
MSUGRA scenario. However it is easier to obtain radiative
electroweak symmetry breaking [REWSB] in large $\tan \beta$ regime
when Higgs masses are split ($m_{H_u} < m_{H_d}$). Another
alternative would be to consider D-term contribution to scalar
masses from breaking the rank of $SO(10)$: $m^2_{(H_u,\ H_d)} =
m^2_{10} \mp 2 D_X, \; m^2_Q = m^2_E = m^2_U = m^2_{16} + D_X, \;
m^2_D = m^2_L = m^2_{16} - 3 D_X$. In this case REWSB is possible
if $D_X > 0$ and the Yukawa unification can be achieved only 
for $\mu < 0$.\cite{Baer_etal}

The Yukawa unification is easy to satisfy with $\mu < 0$, because reasonable 
fits to $b$ quark mass prefer negative
SUSY threshold corrections. In most region of SUSY parameter space these are
dominated by gluino contribution which are negative for negative $\mu$.
However, the discrepancy between SM and experimental values of 
anomalous   
magnetic moment of muon can be explained by SUSY only if 
$\mu > 0$.\cite{g-2}
Also $b\rightarrow s\gamma$ for large $\tan \beta$ strongly prefers $\mu >
0$.\cite{WdB} 

In this analysis we are looking for 
the region of SUSY parameter space with $\mu > 0$ consistent with Yukawa 
unification. 
We start with 11 input parameters at the GUT scale:
$M_G, \, \alpha_G, \, \epsilon_3$, ($\epsilon_3$ is a correction to 
$\alpha_3$ at the GUT scale) which are fixed by gauge coupling unification;
$m_{H_u}, \, m_{H_d}$ which are constrained by requiring correct REWSB;
$\lambda, \, \tan \beta$ which are constrained by Yukawa unification;
and remaining parameters $\mu, \, M_{1/2}, \, m_{16}, \, A$ affect Yukawa 
unification through RG running and weak scale threshold corrections. We scan 
through these parameters to find the region where Yukawa unification can be 
satisfied. 
%We use two (one) loop RG running for dimensionless (dimensionful) 
%parameters from $M_G$ to $M_Z$; include complete one loop SUSY threshold 
%corrections at weak scale; and three loop QCD and one loop QED RG running 
%below $M_Z$. 
The $\chi^2$ function includes 9 observables: 6 precision 
electroweak data $\alpha_{EM},\; G_\mu, \;  \alpha_s(M_Z),\; M_Z, \; M_W, \; 
\rho_{NEW}$ and the 3 fermion masses $M_{top},\;  m_b(m_b), \; M_\tau$.

\begin{figure}[t]
\begin{center}

\input{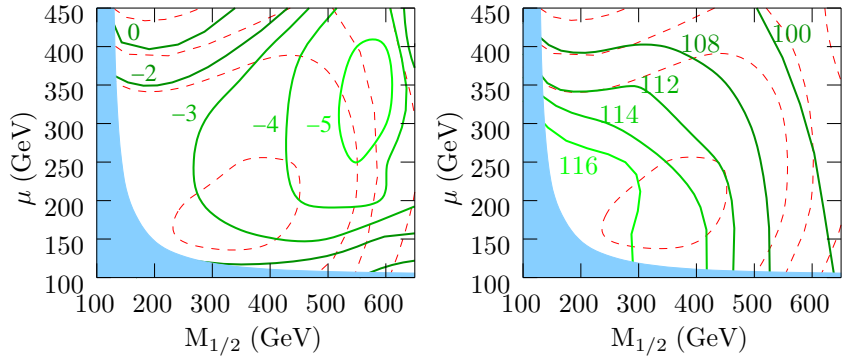}

\caption{
Contours of constant $\Delta m_b$ in \% (left) and $m_{h^0}$ (GeV) (right) for 
$m_{16} = 2000$ GeV with the constant $\chi^2$ contours overlayed (dashed 
lines). 
These correspond to $\chi^2 = 1, \, 2, \, 3 \rm \, and \, 5$ in this order from 
the center of 
the picture out. The shaded region is excluded by chargino mass limit 
$m_{\tilde \chi^\pm} > 103$ GeV.}

\label{figure:1}

\end{center}
\end{figure}

Fig.~\ref{figure:1} shows contours of constant $\chi^2$ (dashed 
lines) for 
$m_{16} = 2000$ GeV. These contours basically correspond to contours of 
constant $m_b$ (not shown in the Fig.~\ref{figure:1})\cite{talks} 
which typically gives 
the largest contribution to $\chi^2$. The contour of $\chi^2 = 1$ 
roughly corresponds to $m_b(m_b) \sim 4.25$ GeV and $\chi^2 = 3$ to 
$m_b(m_b) \sim 4.4$ GeV. Fig.~\ref{figure:1} (left) also shows contours 
of constant SUSY threshold corrections to $m_b$. We see that the acceptable 
fits require $\Delta~m_b~\leq~-~2~\%$.

We find acceptable fits ($\chi^2 < 3$) for $m_{16} > 1200$ GeV, 
$A \sim -2 m_{16}$, $\mu, \, M_{1/2} \ll m_{16}$ and 
$\tan \beta \sim 50 \pm 2$.
Why does Yukawa unification only work in this narrow region of SUSY 
parameter space? How is it possible to get negative threshold corrections 
with positive $\mu$?  The dominant corrections to $m_b$ comes from gluino 
and chargino diagrams. These are proportional to $\tan \beta$ and can 
naturally be as large as 40 \%:
\begin{equation} 
\Delta m_b^{\tilde g} \approx  \frac{2 \alpha_3}{3 \pi}
\frac{\mu m_{\tilde g}}{m_{\tilde b}^2} \tan\beta \;\;\;\;\;\;\;\; 
\rm and \;\;\;\;\;\;\;\;
\Delta m_b^{\tilde \chi^\pm} \approx \frac{\lambda_t^2}{16 \pi^2}
\frac{\mu A_t}{m_{\tilde t}^2} \tan\beta. 
\end{equation}
Gluino corrections are positive for $\mu > 0$ while chargino corrections are 
negative, since $A_t$ runs to an infrared fixed point $\propto - M_{1/2}$.
Hence in order to have a negative total correction to $m_b$, chargino 
corrections must dominate. This can be achieved if $A_t << 0$ and $m_{\tilde 
t} < m_{\tilde b}$. 
Large values of $m_{16}$ are necessary to obtain large enough splitting between
stop and sbottom.
We typically find $m_{\tilde b_1} \sim 3 \, m_{\tilde t_1}$.  This also explains 
why D-term 
splitting scenario with $D_X > 0$ is disfavoured. It makes the
sbottom lighter then the stop already at the GUT scale and leads to 
$m_{\tilde b_1} \le m_{\tilde t_1}$.

The typical spectrum for this region of SUSY parameter space is: quite light 
stop
$\tilde t_1 \sim 150 - 250$ GeV, chargino $\tilde \chi^\pm \sim 100 - 
250$ GeV and scalars of the first two generations above 1 TeV.
Finally in Fig.~\ref{figure:1} (right) we see that Yukawa unification 
prefers 
light Higgs mass in a narrow range, 110 - 118 GeV. 

The range of SUSY parameters with $m_{16} > 1200$ GeV and $M_{1/2} \ll m_{16}$ 
is also preferred by nucleon decay experiments.\cite{DMR}
However, large values of $m_{16} \ge 1200$ GeV lead to very small values of
$a^{NEW}_\mu \le 16 \times 10^{-10}$. 

%\section*{Acknowledgments}
%I would like to thank my collaborators T. Bla\v zek and S. Raby,
%and organisers of the SUSY'01. This work was supported in part by
%DOE grant DOE/ER/01545-???.

\end{document}